\documentclass[12pt]{article}
\usepackage{epsf}
\usepackage{graphicx}

\begin{document}
\title
{Galaxy number counts in a presence of the graviton background}
\author
{Michael A. Ivanov \\
Physics Dept.,\\
Belarus State University of Informatics and Radioelectronics, \\
6 P. Brovka Street,  BY 220027, Minsk, Republic of Belarus.\\
E-mail: ivanovma@gw.bsuir.unibel.by.}

\maketitle

\begin{abstract}In the model of low-energy quantum gravity by the
author, cosmological redshifts are caused by interactions of
photons with gravitons. Non-forehead collisions with gravitons
will lead to an additional relaxation of any photonic flux. Using
only the luminosity distance and a geometrical one as functions of
a redshift in this model, theoretical predictions for galaxy
number counts are considered here. The Schechter luminosity
function with $\alpha =-2.43$ is used. The considered model
provides a good fit to galaxy observations by Yasuda et al. (AJ,
122 (2001) 1104) if the same K-corrections are added. It is shown
that observations of $N(z)$ for different magnitudes $m$ are a lot
more informative than the ones of $N(m).$
\end{abstract}
\section[1]{Introduction }
The standard cosmological model explains observations only under
the circumstance that almost all matter and energy of the Universe
are hidden in some unknown {\it dark} forms. In my model of
low-energy quantum gravity based on the idea of an existence of
the background of super-strong interacting gravitons (for more
details, see \cite{500}), a cosmological redshift is caused by
interactions of photons with gravitons. Non-forehead collisions
with gravitons lead to a very specific additional relaxation of
any photonic flux that gives a possibility of another
interpretation of supernovae 1a data - without any kinematics or
dark energy \cite{500}. I would like to summarize here the main
cosmologically essential consequences of this model. Average
energy losses of a photon with an energy $E $ on a way $dr$
through the graviton background will be equal to: $dE=-aE dr,$
where $a=H/c,$ $H$ is the Hubble constant. If we introduce a new
dimensional constant $D$, so that: $\sigma (E,\epsilon)= D \cdot E
\cdot \epsilon,$ $\sigma (E,\epsilon)$ is a cross-section of
interaction by forehead collisions of a photon with an energy $E$
and a graviton with an energy $\epsilon,$ then we can compute the
Hubble constant in this approach: $H= (1/2\pi) D \cdot \bar
\epsilon \cdot (\sigma T^{4}),$ where $\bar \epsilon$ is an
average graviton energy, and $T$ is a temperature of the
background. The constant $D$ should have the value: $D=0.795 \cdot
10^{-27}{m^{2} / eV^{2}};$ the one may be found from the Newtonian
limit of gravity. If $r$ is a geometrical distance from a source,
then we have for $r(z)$, $z$ is a redshift: $r(z)= ln (1+z)/a.$
None-forehead collisions of photons with gravitons of the
background will lead to a scatter of photons and to an additional
relaxation of a photonic flux, so that the luminosity distance
$D_{L}$ is equal in this approach to: $D_{L}=a^{-1} \ln(1+z)\cdot
(1+z)^{(1+b)/2} \equiv a^{-1}f_{1}(z),$ where $f_{1}(z)\equiv
\ln(1+z)\cdot (1+z)^{(1+b)/2}$ is the luminosity distance in units
of $c/H.$ This luminosity distance function fits supernova
observations very well for roughly $z < 0.5$. It excludes a need
of any dark energy to explain supernovae dimming.
\par
In this paper, I consider galaxy number counts/redshift and
counts/magnitude relations on a basis of this model. I assume here
that a space is flat and the Universe is not expanding.
\section[2]{The galaxy number counts-redshift relation }
Total galaxy number counts $dN(r)$ for a volume element
$dV=d\Omega r^{2}dr$ is equal to: $dN(r)=n_{g}dV=n_{g}d\Omega
r^{2}dr,$ where $n_{g}$ is a galaxy number density (it is constant
in the no-evolution scenario), $d\Omega$ is a solid angle element.
Using the function $r(z)$ of this model, we can re-write  galaxy
number counts as a function of a redshift $z$:
\begin{equation}dN(z)=n_{g}d\Omega (H/c)^{-3} {ln^{2}(1+z) \over
{1+z}}dz. \end{equation}
Let us introduce a function (see \cite{123}) $$f_{2}(z)\equiv
{(H/c)^{3}dN(z) \over n_{g}d\Omega z^{2} dz};$$ then we have for
it in this model:
\begin{equation}f_{2}(z)= {ln^{2}(1+z) \over {z^{2}(1+z)}}.\end{equation}
A graph of this function is shown in Fig. 1; the typical error bar
and data point are added here from paper \cite{72} by Loh and
Spillar. There is not a visible contradiction with observations.
{\it There is not any free  parameter in the model to fit this
curve;} it is a very rigid case.
\begin{figure}[th]
\epsfxsize=12.98cm \centerline{\epsfbox{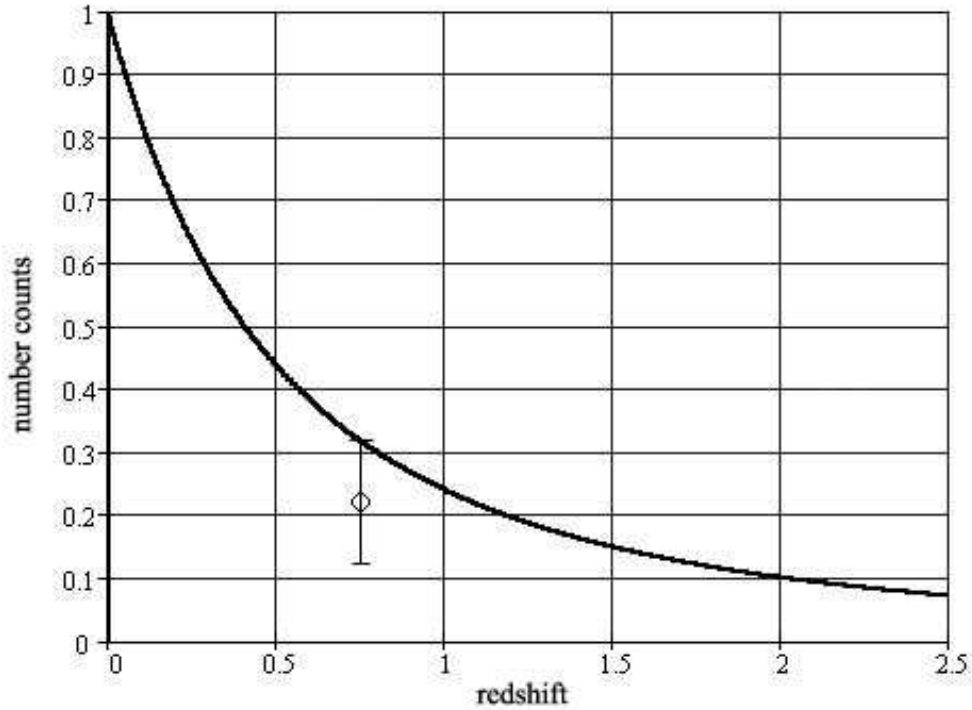}} \caption{Number
counts $f_{2}$ as a function of the redshift in this model. The
typical error bar and data point are taken from paper \cite{72} by
Loh and Spillar.}
\end{figure}
\par It is impossible to count a {\it total} galaxy number for big
redshifts so as very faint galaxies are not observable. For
objects with a fixed luminosity, it is easy to find how their
magnitude $m$ changes with a redshift. So as $dm(z)$ under a
constant luminosity is equal to: $dm(z)=5 d(lg D_{L}(z)),$ we have
for $\Delta m(z_{1},z_{2})\equiv \int_{z_{1}}^{z_{2}} dm(z):$
\begin{equation}\Delta m(z_{1},z_{2})=5 lg(f_{1}(z_{2})/f_{1}(z_{1})).
\end{equation}
This function is shown in Fig.2 for $z_{1}=0.001; 0.01; 0.1.$
\begin{figure}[th]
\epsfxsize=12.98cm \centerline{\epsfbox{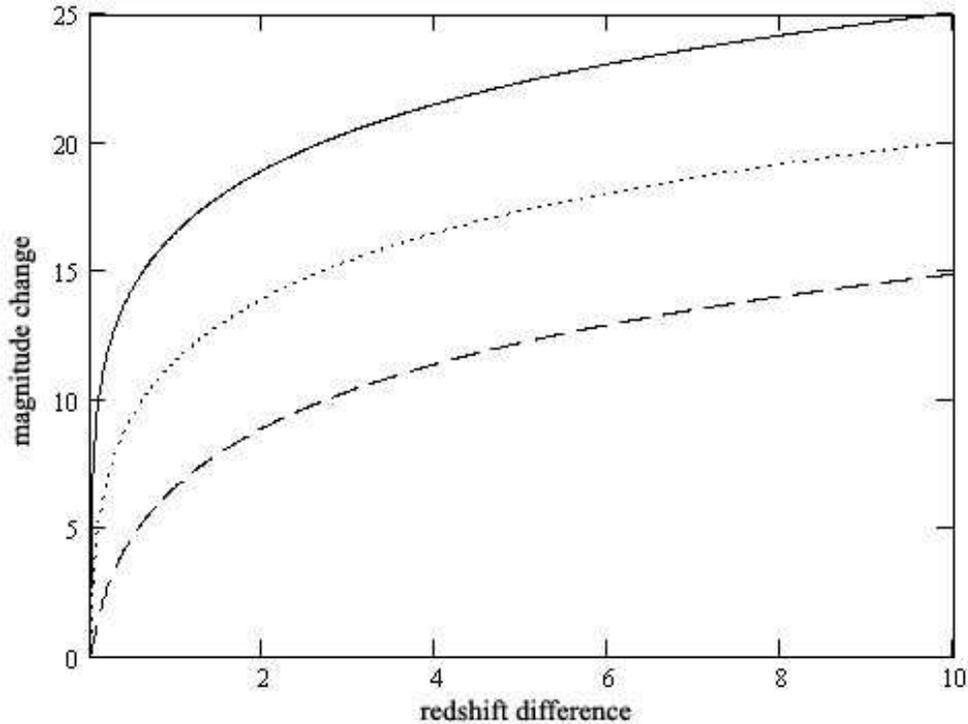}}
\caption{Magnitude changes $\Delta m$  as a function of the
redshift difference $z_{2}-z_{1}$ in this model for $z_{1}=0.001$
(solid); 0.01 (dot); 0.1 (dash).}
\end{figure}
\par I would like to note that a very fast {\it initial} growth of
the luminosity distance with a redshift $z$ in this model might
explain the observed excess of faint blue galaxy number counts
above an expected one in the standard model (for example, see
\cite{122}). A galaxy color depends on a redshift, and a galaxy
dimming depends on the luminosity distance, because by big values
of the ratio $\Delta m(z_{1},z_{2})/(z_{2}-z_{1})$ in a region of
small redshifts and by a further much slower change of it (see
Fig.3) an observer will see many faint but blue enough galaxies in
this region (in the no-evolution scenario).
\begin{figure}[th]
\epsfxsize=12.98cm \centerline{\epsfbox{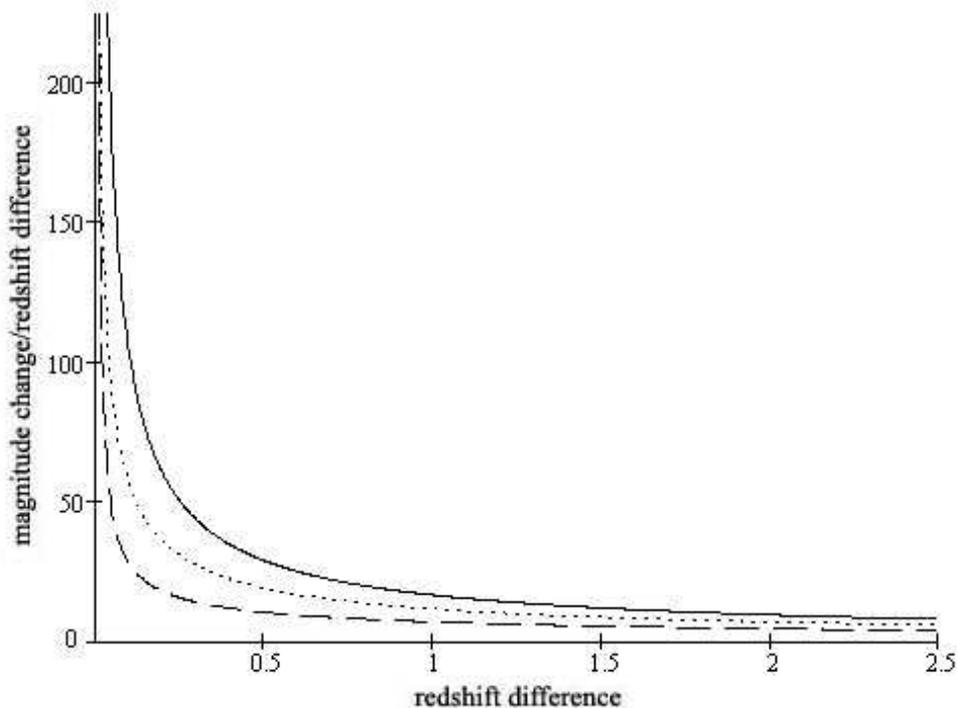}} \caption{To a
possible explanation of the excess of faint blue galaxy number
counts: $\Delta m(z_{1},z_{2})/(z_{2}-z_{1})$ vs. the redshift
difference $z_{2}-z_{1}$ in this model for $z_{1}=0.001$ (solid);
0.01 (dot); 0.1 (dash).}
\end{figure}
\section[3]{Taking into account the galaxy luminosity function}
Galaxies have different luminosities  $L,$ and we can write
$n_{g}$ as an integral: $n_{g}=\int dn_{g}(L),$ where
$dn_{g}(L)=\eta (L)dL,$ $\eta (L)$ is the galaxy luminosity
function. I shall use here the Schechter luminosity function
\cite{73}:
\begin{equation}
\eta (L)dL=\phi_{\ast}({L \over L_{\ast}})^{\alpha} exp(-{L \over
L_{\ast}}) d({L \over L_{\ast}})\end{equation} with the parameters
$\phi_{\ast},$ $L_{\ast},$ $\alpha.$\footnote{To turn aside the
problem with divergencies of this function by small $L$ for
negative values of $\alpha,$  all computations are performed here
for $z>0.001.$} So as we have by a definition of the luminosity
distance $D_{L}(z)$ that a light flux $I$ is equal to: $I={L \over
4\pi D_{L}^{2}(z)},$ and a visible magnitude $m$ of an object is
$m=-2.5 \lg I + C,$ where $C$ is a constant, then $m$ is equal to:
\begin{equation}m=-2.5 \lg I +5 \lg D_{L}(z)
+(C-4\pi).\end{equation} We can write for $L:$
\begin{equation} L=A\cdot{D_{L}^{2}(z) \over \kappa^{m}},\end{equation}
where $\kappa=10^{0.4}, \ A=const. $ For a thin layer with
$z=const$ we have:
$$ dL={\partial L \over \partial m}\cdot dm,$$ where
\begin{equation} {\partial L \over \partial m}=- m \kappa \cdot A {D_{L}^{2}(z)
\over \kappa^{m}}= -m\kappa L.\end{equation} Then
\begin{equation}
dn_{g}(m,z)=-(\phi_{\ast}\kappa)\cdot l^{\alpha}(m,z)
\exp(-l(m,z))\cdot (m\cdot l(m,z))dm,
\end{equation}
where $(-dm)$ corresponds to decreasing $m$ by growing $L$ when
$z=const,$ and $$l(m,z)\equiv {L(m,z) \over L_{\ast}}.$$
\par Let us introduce a function $f_{3}(m,z)$ with a differential
\begin{equation}
df_{3}(m,z)\equiv {dN(m,z) \over d\Omega (-dm).}
\end{equation}
We have for this differential in the model:
\begin{equation}
df_{3}(m,z)=({\phi_{\ast}\kappa \over a^3})\cdot m\cdot
l^{\alpha+1}(m,z)\cdot \exp(-l(m,z))\cdot {ln^{2}(1+z) \over
(1+z)}dz,
\end{equation}
where $a=H/c$, $H$ is the Hubble constant. An integral on $z$
gives the galaxy number counts/magnitude relation:
\begin{equation}
f_{3}(m)=({\phi_{\ast}\kappa \over a^3})\cdot m\cdot
\int_{0}^{z_{max}} l^{\alpha+1}(m,z)\cdot \exp(-l(m,z))\cdot
{ln^{2}(1+z) \over (1+z)}dz;
\end{equation}
I use here an upper limit $z_{max} =10.$ To compare this function
with observations by Yasuda et al. \cite{77}, let us choose the
normalizing factor from the condition: $f_{3}(16)=a(16), $ where
\begin{equation}
a(m)\equiv A_{\lambda}\cdot 10^{0.6(m-16)}
\end{equation}
is the function assuming "Euclidean" geometry and giving the best
fit to observations \cite{77}, $A_{\lambda}=const$ depends on the
spectral band. In this case, we have two free parameters -
$\alpha$ and $L_{\ast}$ - to fit observations, and the latter one
is connected with a constant $A_{1}\equiv {A \over a^{2}L_{\ast}}$
if $$l(m,z)=A_{1}{f_{1}^{2}(z) \over \kappa^{m}}.$$
\par If we use the magnitude scale in which $m=0$ for Vega then
$C=2.5\lg I_{Vega},$ and we get for $A_{1}$ by $H=2.14\cdot
10^{-18}\ s^{-1}$ (it is a theoretical estimate of $H$ in this
model \cite{500}):
\begin{equation}
A_{1}\simeq 5\cdot 10^{17}\cdot {L_{\odot} \over L_{\ast}},
\end{equation}
where $L_{\odot}$ is the Sun luminosity; the following values are
used: $L_{Vega}=50 L_{\odot},$ the distance to Vega $r_{Vega}=26 \
LY.$
\par Without the factor $m,$ the function $f_{3}(m)$ by $\exp(-l(m,z)\rightarrow
1$ would be close to $a(m)$ by $\alpha =-2.5.$ Matching values of
$\alpha$ shows that $f_{3}(m)$ is the closest to $a(m)$ in the
range $10<m<20$ by $\alpha =-2.43.$  The ratio ${{f_{3}(m)
-a(m)}\over a(m)}$ is shown in Fig.4 for different values of
$A_{1}$ by this value of $\alpha$. All such the curves conflow by
\begin{figure}[th]
\epsfxsize=12.98cm \centerline{\epsfbox{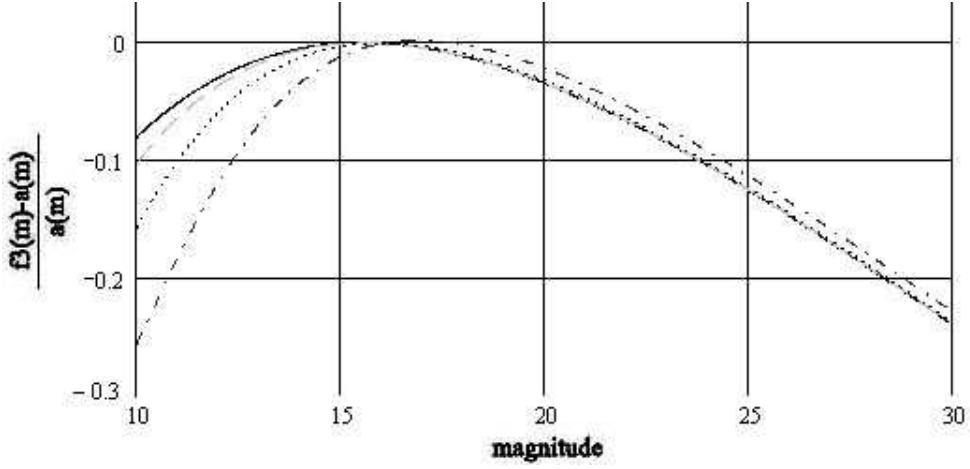}}
\caption{The relative difference $(f_{3}(m)-a(m))/a(m)$ as a
function of the magnitude $m$ for $\alpha=-2.43$ by
$10^{-2}<A_{1}<10^{2}$ (solid), $A_{1}=10^{4}$ (dash),
$A_{1}=10^{5}$ (dot), $A_{1}=10^{6}$ (dadot). }
\end{figure}
$A_{1}\leq 10^2$ (or $5\cdot 10^{15}<L_{\ast}$), i.e. observations
of the galaxy number counts/magnitude relation are {\it
non-sensitive} to $A_{1}$ in this range. For fainter magnitudes
$20<m<30$, the behavior of all curves is identical: they go below
of the ratio value $1$ with the same slope. If we compare this
figure with Figs. 6,10,12 from \cite{77}, we see that the
considered model provides a no-worse fit to observations than the
function $a(m)$ if the same K-corrections are added (I think that
even a better one if one takes into account positions of
observational points in Figs. 6,10,12 from \cite{77} by $m<16$ and
$m>16$) for the range $10^2<A_{1}< 10^7$ that corresponds to
$5\cdot 10^{15}>L_{\ast}> 5\cdot 10^{10}.$
\par Observations of $N(z)$ for different magnitudes are a lot more
informative. If we define a function $f_{4}(m,z)$ as
\begin{equation}
f_{4}(m,z)\equiv ({a^3 \over \phi_{\ast}\kappa})\cdot {df_{3}(m,z)
\over dz},
\end{equation}
this function is equal in the model to:
\begin{equation}
f_{4}(m,z)= m\cdot l^{\alpha+1}(m,z)\cdot \exp(-l(m,z))\cdot
{ln^{2}(1+z) \over (1+z)}.
\end{equation}
\par Galaxy number counts in the range $m_{1}<m<m_{2}$ are
proportional to the function:
\begin{equation}
f_{5}(m_{1},m_{2})\equiv \int_{m_{1}}^{m_{2}} f_{4}(m,z)dm=
\end{equation}
$$=
 \int_{m_{1}}^{m_{2}} m\cdot l^{\alpha+1}(m,z)\cdot
\exp(-l(m,z))\cdot {ln^{2}(1+z) \over (1+z)}dm.$$
\begin{figure}[th]
\epsfxsize=10.98cm\centerline{\epsfbox{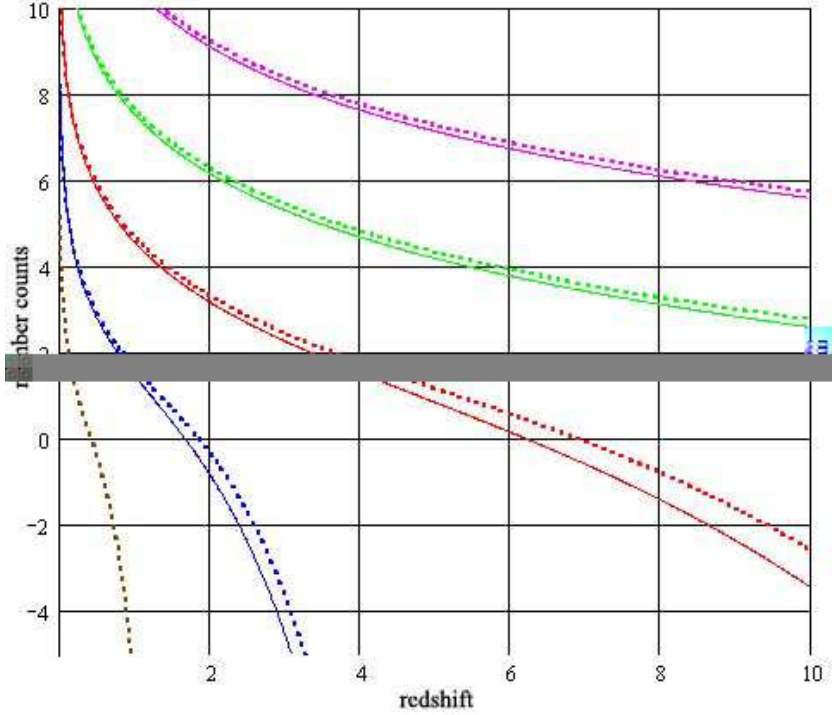}}
\caption{Number counts $f_{4}(m,z)$ (dot) and $f_{5}(m_{1},m_{2})$
(solid) (logarithmic scale) as a function of the redshift by
$A_{1}=10^{5}$ for $\alpha=-2.43,\ m_{1}=10$ and  different values
of $m=m_{2}:$ 15 (blue), 20 (red), 25 (green), and 30 (mag);
$m=10$ (brown, only $f_{4}(m,z)$),.}
\end{figure}
Graphs of both $f_{4}(m,z)$ and $f_{5}(m_{1},m_{2})$ are shown in
Fig. 5 by $\alpha =-2.43,\ A_{1}=10^{5} $; they are very similar
between themselves. We see that even the observational fact that a
number of visible galaxies by $z\sim 10$ is very small allows us
to restrict a value of the parameter $A_{1}$ much stronger than
observations of $N(m)$.
\section[4]{Quasar number counts}
For quasars, we can attempt to compute the galaxy number
counts/redshift relation using  Eq. 16 with another luminosity
function $\eta{'} (l(m,z))$:
\begin{equation}
f_{5}(m_{1},m_{2})\equiv \int_{m_{1}}^{m_{2}} f_{4}{'}(m,z)dm=
 \int_{m_{1}}^{m_{2}} m\cdot l(m,z)\cdot \eta{'} (l(m,z))\cdot
{ln^{2}(1+z) \over (1+z)}dz.
\end{equation}
The following luminosity functions were probed here (see Fig. 6):
the Schechter one with $\alpha =0$, $A_{1}=10^{6.55}$ (blue); the
double power law \cite{74,76}:
\begin{equation}
\eta{'} (l(m,z)) \propto {1 \over l^{-\alpha}(m,z) +
l^{-\beta}(m,z)}
\end{equation}
with $\alpha =-3.9,\ \beta=1.6, \ A_{1}=4.5\cdot 10^{6}$ (green);
the Gaussian one:
\begin{equation}
\eta{'} (l(m,z)) \propto \exp({-(l(m,z)-1)^{2} \over 2\sigma^2})
\end{equation}
with $\sigma =0.5,\  A_{1}=4.5\cdot 10^{6}$ (brown, dot); the
combined one:
\begin{equation}
\eta{'} (l(m,z)) \propto l^{\alpha}(m,z)\cdot
\exp({-(l(m,z)-1)^{2} \over 2\sigma^2})
\end{equation}
with two sets of parameters: $\alpha =-1.45, \ \sigma =0.6, \
A_{1}=1.3\cdot 10^{6}$ (red, solid) and $\alpha =-1.4, \ \sigma
=0.7, \ A_{1}=3\cdot 10^{6}$ (red, dot). There is a couple of
\begin{figure}[th]
\epsfxsize=12.98cm
\centerline{\epsfbox{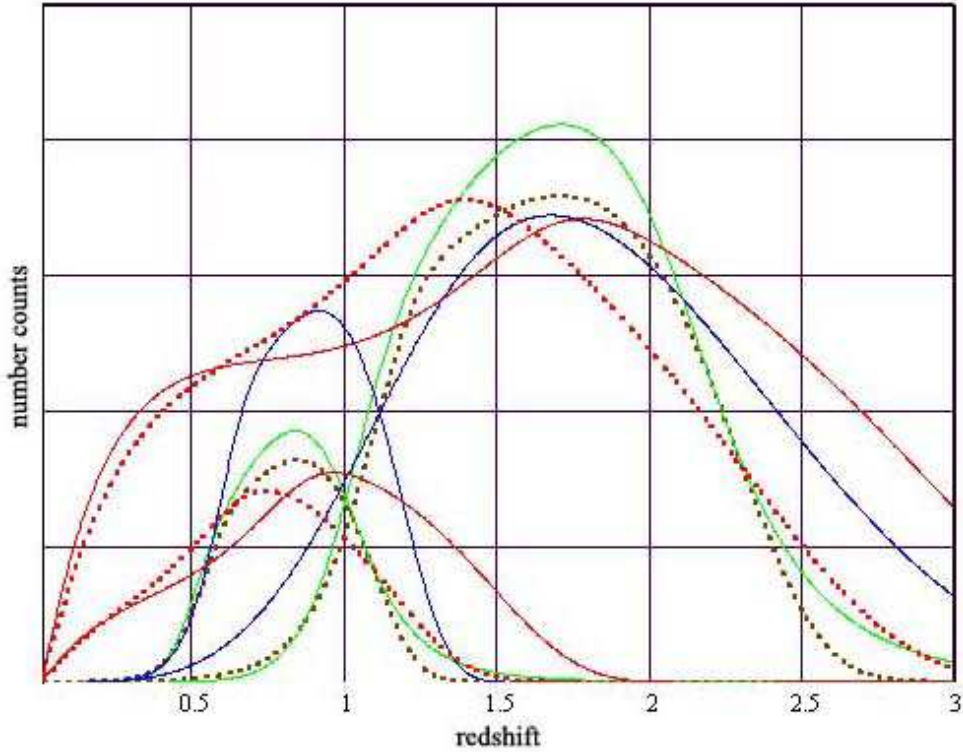}}
\caption{QSO number counts $f_{5}(m,z)$ (arbitrary units) as a
function of the redshift for different luminosity functions:
Gaussian (brown, dot), the double power law (green), Schechter's
(blue), combined (red, solid and dot) with parameters given in the
text. The left-shifted curve of each couple corresponds to the
range $16<m<18.25,$ another one corresponds to $18.25<m<20.85.$}
\end{figure}
curves for each case: the left-shifted curve of any couple
corresponds to the range $16<m<18.25,$ another one corresponds to
$18.25<m<20.85.$ These ranges are chosen the same as in the paper
by Croom et al. \cite{74}, and you may compare this figure with
Fig. 3 in \cite{74}. We can see that the theoretical distributions
reflect only some features of the observed ones but not an entire
picture. In all these cases, a slope of an analog of
$\log(f_{3}(m))$ near $m=18$ is in the range 0.29 - 0.325, when
quasar observations give a larger slope (see Fig. 4, 21 in
\cite{74} and Fig. 13 in \cite{75}; in the latter paper, this
slope has been evaluated to be equal to about 1). We can summarize
that, as well as in the standard cosmological model, it is
impossible to fit quasar observations using some simple luminosity
function with fixed parameters.

\par In the standard model, an easy way exists to turn aside this
difficulty: one ascribes it to a quasar "evolution", then a
luminosity function (for example, the double power law
\cite{74,76}) is modified for different redshifts to take into
account this "evolution". There exist two manner to do it: one may
consider $L_{\ast}$ as a function of a redshift (pure luminosity
evolution) \cite{74} or one may assume that indices $\alpha$ and
$\beta$ of the distribution (double power law) vary with $z$
\cite{76} - in both variants, it is possible to fit observations
in some range of redshifts; of course, there are many other
descriptions of the "evolution" \cite{75}. It is strange only that
"evolutions" are not concerted: we can see exponential, quadratic
and other kinds of them - and it means that there is not any real
evolution: we deal with a pure fine art of fitting, nothing more.
In the considered model, this way is forbidden.
\par I think that it is necessary to consider some theoretical
model of a quasar activity to get a distribution of
"instantaneous" luminosities. It is known that the typical
lifetime of individual quasars is uncertain by several orders of
magnitude; a lifetime of  $4 \cdot 10^7$ years may be considered
as an average value \cite{771}. If one considers a quasar light
curve $L(t)$ (in a manner which is similar to the one by Hopkins
et al. \cite{78}) in a parametric form, it is possible to get the
luminosity function which takes into account a probability to
observe a quasar with a given luminosity. Let us consider the two
simple examples. The simplest case is a constant luminosity $L$ of
any quasar during its lifetime $\tau$. If initial moments of
quasar activity are distributed uniformly in time and may be
described by a frequency $\nu$, then a probability $P_{obs}$ to
observe a quasar will be equal to:
\begin{equation}
P_{obs}=\int_{0}^{\tau}\exp(-\nu t{'})d(\nu
t{'})=1-\exp(-\nu\tau).
\end{equation}
For $\nu\tau\ll 1$ we have $P_{obs}\simeq \nu\tau.$ If we further
assume that $\tau\propto 1/L,$ i.e. that a full emitted quasar
energy is constant, then a distribution of observable luminosities
is
\begin{equation}
\eta{'}(L)\propto \eta(L)\cdot 1/L,
\end{equation}
where $\eta(L)$ is an initial distribution of values of $L$.
\par The second example is the quasar exponential light curve:
\begin{equation}
L(t)=L_{0}\exp(-t/\tau),
\end{equation}
where $\tau$ is a lifetime, $L_{0}$ is an initial luminosity. If
$L_{0}$ has a distribution $\eta(L_{0})$, then we get:
\begin{equation}
\eta{'}(L) \propto \int
\eta(L_{0})\cdot[\exp(t_{max}/\tau(L_{0}))-1]^{-1}\cdot
(L_{0}/L)^{2-\nu\tau(L_{0})}\cdot dL_{0},
\end{equation}
where $t_{max}$ is a maximum time during which one can distinguish
a quasar from a host galaxy, and $\tau$ depends on $L_{0}$ in some
manner. We see that even in this simple toy example the dependence
on $\tau$ is not trivial.
\par In a general case, it is necessary to describe both - front
and back - slopes of a quasar light curve. Together with a total
emitted energy (or a peak luminosity), we need at least three
independent parameters; if we take into account their random
distributions, this number should be at least doubled.

\section[5]{Conclusion}
Starting from a micro level and considering interactions of
photons with single gravitons, we can find the luminosity distance
and a geometrical distance in this approach. Using only these
quantities, I compute here galaxy number counts-redshift and
galaxy number counts-magnitude relations for a case of a flat
non-expanding universe. It has been shown here that they are in a
good accordance with observations. It may be important as for
cosmology as for a theory of gravity.


\begin{thebibliography}{References                        }
\bibitem{500}
Ivanov, M.A. Gravitons as super-strong interacting particles, and
low-energy quantum gravity. In the book {\it Focus on Quantum
Gravity Research,} Nova Science, 2006, Chapter 3 (in press);
[hep-th/0506189 v3]; [http://ivanovma.narod.ru/nova04.pdf].
\bibitem{123}
Cunha, J. V., Lima, J. A. S., Pires, N. {\it Astronomy and
Astrophysics} 2002, {\it 390,} 809.
\bibitem{72}
Loh, E.D. and Spillar, E.J. {\it ApJ} 1986,
{\it 307,} L1.
\bibitem{122}
Driver, S.P., et al. {\it ApJ} 1998, {\it 496,} L93.
\bibitem{73}
Schechter, P.L. {\it ApJ} 1976, {\it 203,} 297.
\bibitem{77}
Yasuda, N. et al. {\it Astron.J.} 2001, {\it 122,} 1104.
\bibitem{74}
Croom, S.M. et al. {\it MNRAS} 2004, {\it 349,} 1397.
\bibitem{76}
Hopkins, P.F. et al. [astro-ph/0605678].
\bibitem{75}
Richards, G.T. et al. [astro-ph/0601434].
\bibitem{771}
Martini, P. and Weinberg, D.H. {\it ApJ} 2001, {\it 547,} 12
[astro-ph/0002384].
\bibitem{78}
Hopkins, P.F. et al. {\it ApJ} 2006, {\it 632,} 700
[astro-ph/0508299].
\end{thebibliography}
\end{document}